\documentstyle[11pt,newpasp,twoside,epsf]{article}

\reversemarginpar

\begin{document}

\title{Close Binaries with Two Compact Objects}
\author{V. Kalogera}
\affil{Harvard-Smithsonian Center for Astrophysics, 60 Garden St.,
Cambridge, MA 02138, USA}

 \begin{abstract}
 The coalescence of close binary systems with two compact objects (neutron
stars and black holes) are considered to be promising sources of
gravitational waves for the currently built laser interferometers. Here, I
review the current Galactic coalescence estimates derived both
theoretically and empirically. I discuss the uncertainties involved as
well as ways of obtaining an upper limit to the coalescence rate of two
neutron stars. 
 \end{abstract}

\section{Introduction}

The late inspiral phase due to gravitational radiation of close binaries
with two compact objects, neutron stars (NS) or black holes (BH), is
expected to be a major source of gravitational waves for ground-based
interferometers, such as LIGO and VIRGO). The binary radio pulsar PSR
B1913+16 (Hulse \& Taylor 1975) is the prototypical NS--NS system and has
provided a remarkable empirical confirmation of general relativity with
the measurement of orbital decay due to gravitational radiation (Taylor \&
Weisberg 1982). In addition to NS--NS binaries, the existence of close
BH--NS and BH--BH systems has also been predicted theoretically, and they
are considered as possible sources of gravitational radiation.

The expected detection rate of inspiral events depends crucially on the
rate of such events in the Galaxy and by extrapolation out to the maximum
distance an inspiral event could be detected.  In what follows I review
the current estimates of Galactic coalescence rates, obtained
theoretically and empirically, and I discuss the associated uncertainties.
For the case of NS--NS binaries, I also discuss a new way of obtaining an
empirical upper bound to their coalescence rate. The scale for all these
rates is set by requiring that the detection rate for the ``enhanced''
LIGO is $\sim 2-3$ events per year. Given the amplitude of the signals
(e.g., Thorne 1996), the requirement for the NS--NS Galactic rate is $\sim
10^{-5}$\,yr$^{-1}$ (detected out to $\sim 200$\,kpc) and for the BH--BH
Galactic rate $\sim 2\times 10^{-7}$\,yr$^{-1}$ (detected out to $\sim
700$\,kpc).

\section{Theoretical Estimates}

The formation rate of {\em coalescing} binary compact objects (i.e.,
systems in close enough orbits that coalesce within a Hubble time) can be
calculated, given a sequence of evolutionary stages leading to binary
compact object formation.  Over the years a relatively standard picture
(van den Heuvel 1976) has been formed describing the birth of such systems
based on the consideration of NS--NS binaries (for variations of this
picture see Brown 1995; Terman \& Taam 1995):  The initial binary
progenitor consists of two binary members massive enough to eventually
collapse into a NS or a BH. The evolutionary path involves multiple phases
of stable or unstable mass transfer, common-envelope phases, and accretion
onto compact objects, as well as two core collapse events. Theoretical
modeling of this formation process has been undertaken by various authors
using population synthesis techniques, Such studies provide us with {\em
ab initio} predictions of coalescence rates. The evolution of an ensemble
of primordial binaries with assumed initial properties is followed until a
coalescing binary is formed. The changes in the binary properties at the
end of each evolutionary stage are calculated based on our current
understanding of the processes involved: wind mass loss from massive
hydrogen- and helium-rich stars, mass and angular-momentum losses during
mass transfer phases, dynamically unstable mass transfer and
common-envelope evolution, effects of highly super-Eddington accretion
onto NS, and supernova explosions with kicks imparted to newborn NS or
even BH. Given that several of these phases are not well understood, the
results of population synthesis are expected to depend on the assumptions
made in the treatment of the various processes. Therefore, exhaustive
parameter studies are required by the nature of the problem.

Recent studies of the formation rate of coalescing compact objects
(Lipunov, Postnov, \& Prokhorov 1997; Fryer, Burrows, \& Benz 1998;
Portegies-Zwart \& Yungel'son 1998;  Brown \& Bethe 1998; Fryer, Woosley,
\& Hartmann 1999)  have explored the input parameter space and the
robustness of the results at different levels of (in)completeness. Kicks
imparted to NS and BH at birth are the best studied of all model
assumptions. In the case of NS--NS coalescence the range of predicted
Galactic rates is $<10^{-7}~-~5\times 10^{-4}$\,yr$^{-1}$. This large
range indicates the importance of supernovae (two in this case) in
binaries. Variations in the assumed mass-ratio distribution for the
primordial binaries can {\em further} change the predicted rate by about a
factor of $10$, while assumptions of the common-envelope phase add another
factor of about $10-100$. Variation in other parameters typically affects
the results by factors of two or less. Predicted rates for BH--NS and
BH--BH binaries lie in the ranges $< 10^{-7}~-~10^{-4}$\,yr$^{-1}$ and
$<10^{-7}~-~10^{-5}$\,yr$^{-1}$, respectively when the kick magnitude to
both NS and BH is varied. Other uncertain factors such as the critical
progenitor mass for NS and BH formation lead to variations of the rates by
factors of $10-50$.

Although results from population syntheses regarding binary properties,
such as orbital sizes, eccentricities, center-of-mass velocities, can be
quite robust, it is evident that predicted rates, i.e., the absolute
normalization of the models, cover a wide range of values (typically 3-4
orders of magnitude).  Given these results it seems fair to say that
population synthesis calculations have a rather limited predictive power
and provide fairly loose constraints on coalescence rates.

\section{Empirical Estimates}

The observed sample of coalescing NS--NS binaries (PSR B1913+16 and PSR
B1534+12) provides us with another way of estimating their coalescence
rate. An empirical estimate can be obtained using their observed pulsar
and binary properties along with models of selection effects in radio
pulsar surveys. For each observed object, a scale factor can be calculated
based on the fraction of the Galactic volume within which pulsars with
properties identical to those of the observed pulsar could be detected by
any of the radio pulsar surveys, given their detection thresholds. This
scale factor is a measure of how many more pulsars like the ones detected
in the coalescing NS--NS systems exist in our galaxy. The coalescence rate
can then be calculated based on the scale factors and estimates of
detection lifetimes summed up for all the observed systems.  This basic
method was first used by Phinney (1991) and Narayan, Piran, \& Shemi
(1991)  who estimated the Galactic rate to be $\sim 10^{-6}$\,yr$^{-1}$.

Since then, estimates of the NS--NS coalescence rate have known a
significant downward revision primarily because of (i) the increase of the
Galactic volume covered by radio pulsar surveys with no additional
coalescing NS--NS being discovered (Curran \& Lorimer 1995), (ii) the
increase of the distance estimate for PSR B1534+12 based on measurements
of post-Newtonian parameters (Stairs et al.\ 1998) (iii) revisions of the
lifetime estimates (van den Heuvel \& Lorimer 1996; Arzoumanian, Cordes,
\& Wasserman 1999). In addition, a significant upward correction factor
($\sim 7-10$) has been used recently to account for the faint end of the
radio pulsar luminosity function. The most recently published study
(Arzoumanian et al.\ 1999) gives a lower limit of $2\times
10^{-7}$\,yr$^{-1}$ and a ``best'' estimate of $\sim 6-10\times
10^{-7}$\,yr$^{-1}$, which agrees with other recent estimates of
$2-3\times10^{-6}$\,yr$^{-1}$ (Stairs et al.\ 1998; Evans et al.\ 1999).  
Additional uncertainties arise from estimates of pulsar ages and
distances, the pulsar beaming fraction, the spatial distribution of DNS in
the Galaxy.

Despite all these uncertainties the empirical estimates of the NS--NS
coalescence rate appear to span a range of $\sim 1-2$ orders of
magnitude, which is narrow compared to the range covered by the
theoretical estimates.

\subsection{Small Number Sample and Pulsar Luminosity Function}

One important limitation of empirical estimates of the coalescence rates
is that they are derived based on {\em only two} observed NS--NS systems,
under the assumption that the observed sample is representative of the
true population, particularly in terms of their radio luminosity. Assuming
that DNS pulsars follow the radio luminosity function of young pulsars and
that therefore their true population is dominated in number by
low-luminosity pulsars, it can be shown that the current empirical
estimates most probably underestimate the true coalescence rate.  If a
small-number sample is drawn from a parent population dominated by
low-luminosity (hence hard to detect) objects, it is statistically more
probable that the sample will actually be dominated by objects from the
high-luminosity end of the population. Consequently, the empirical
estimates based on such a sample will tend to overestimate the detection
volume for each observed system, and therefore underestimate the scale
factors and the resulting coalescence rate.

This effect can be clearly demonstrated with a Monte Carlo experiment
(Kalogera et al.\ 1999) using simple models for the pulsar luminosity
function and the survey selection effects. As a first step, the average
observed number of pulsars is calculated given a known ``true'' total
number of pulsars in the Galaxy (thick-solid line in Figure 1a). As a
second step, a large number of sets consisting of ``observed'' (simulated)
pulsars are realized using Monte Carlo methods. These pulsars are drawn
from a Poisson distribution of a given mean number ($<N_{\rm obs}>$) and
have luminosities assigned according to the assumed luminosity function.  
Based on each of these sets, one
can estimate the total number of pulsars in the Galaxy using empirical
scale factors, as is done for the real observed sample. The many
(simulated) `observed' samples can then be used to obtain the distribution
of the estimated total Galactic numbers ($N_{\rm est}$) of pulsars. The
median and 25\% and 75\% percentiles of this distribution are plotted as a
function of the assumed number of systems in the (fake) `observed' samples
in Figure 1a (thin-solid and dashed lines, respectively).

 \begin{figure}
 \plottwo{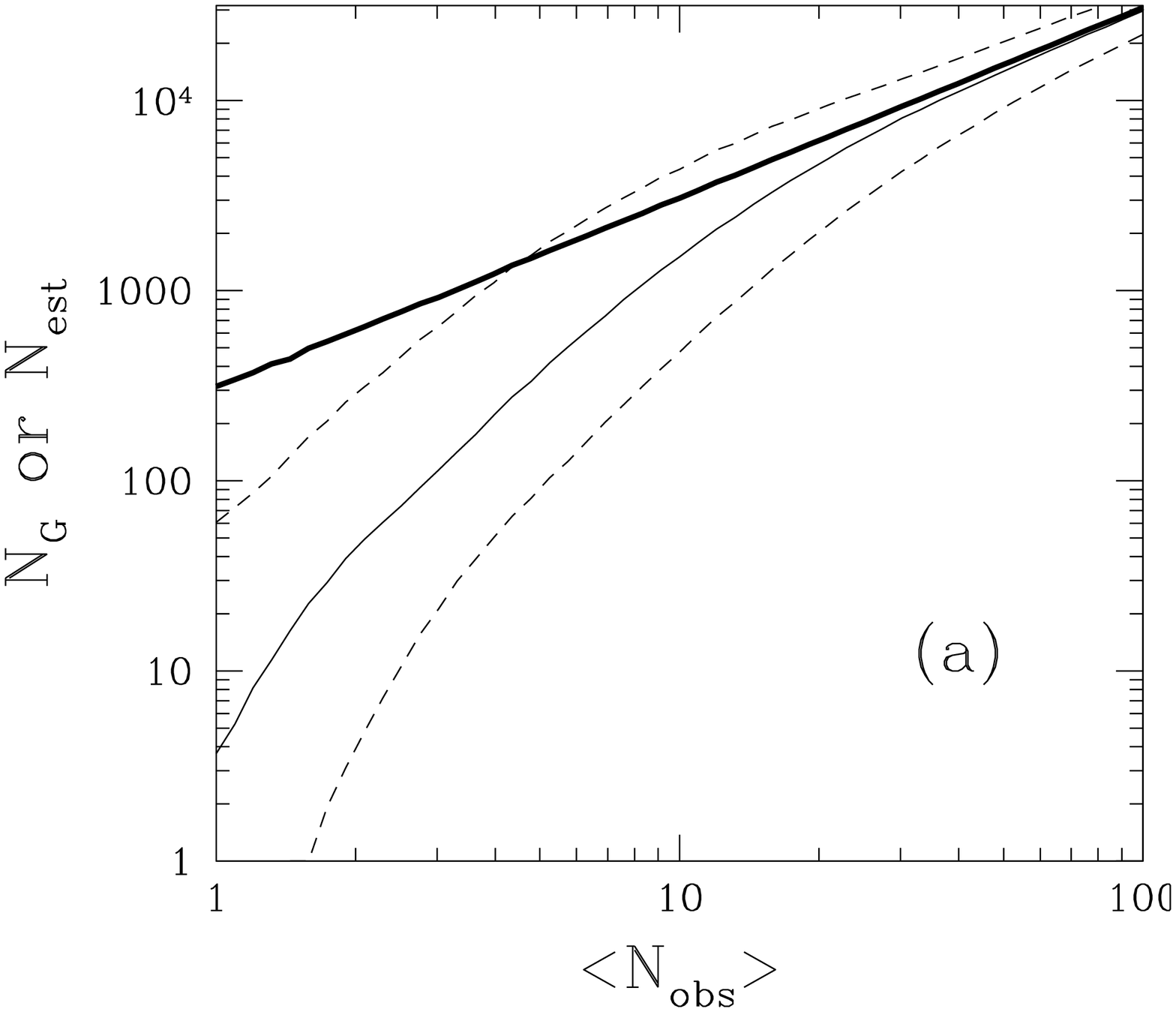}{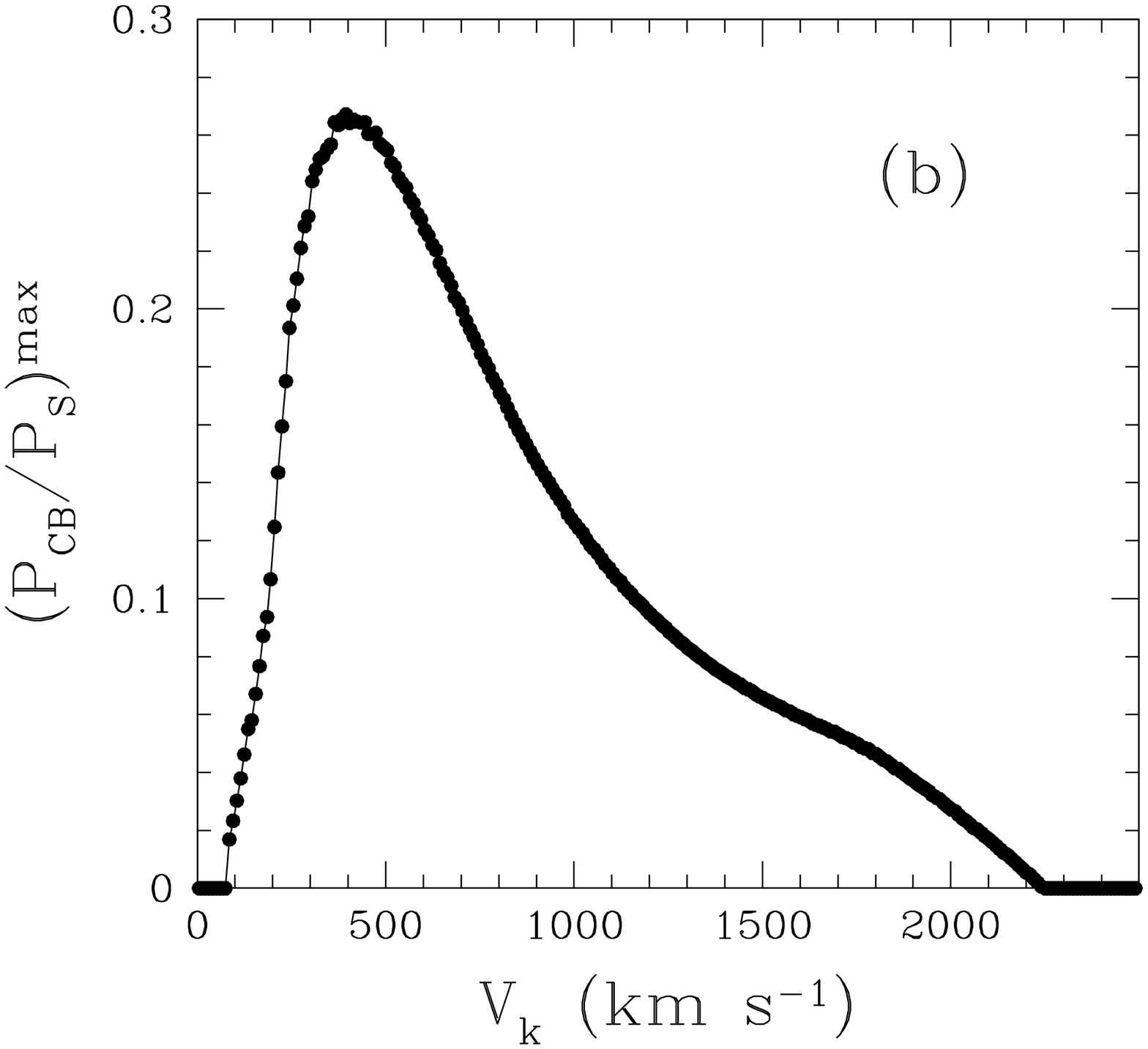}
 \caption{(a) Bias of the empirical estimates of the DNS coalescence rate
because of the small-number observed sample. See text for details. (b)
Maximum probability ratio for the formation of coalescing DNS and the
disruption of binaries as a function of the kick magnitude at the second
supernova.}
 \end{figure}

It is evident that, in the case of small-number observed samples (less
than $\sim 10$ objects), the estimated total number, and hence the
estimated coalescence rate, can be underestimated by a significant factor.
For a two-object sample, for example, the true rate maybe higher by more
than a factor of ten. This correction factor associated with the faint-end
of the luminosity function can be applied to the estimated NS--NS
coalescence rate in place of the factor of $\sim 7-10$ used so far from a
direct extrapolation of the luminosity function.

\section{Limits on the NS--NS Coalescence Rate}

Observations of NS-NS systems and isolated pulsars related to NS--NS
formation allow us to obtain upper limits on their coalescence rate.
Depending on how their value compares to the ``enhanced'' LIGO requirement
such limits can provide us with valuable information about the prospects
of gravitational-wave detection.

Bailes (1996) used the absence of any young pulsars detected in NS--NS
systems and obtained a rough upper limit to the rate of $\sim
10^{-5}$\,yr$^{-1}$, while recently Arzoumanian et al.\ (1999) reexamined
this in more detail and claimed a more robust upper limit of $\sim
10^{-4}$\,yr$^{-1}$.

An upper bound to the NS--NS coalescence rate can also be obtained by
combining our
theoretical understanding of orbital dynamics (for supernovae with NS
kicks in binaries)  with empirical estimates of the birth rates of {\em
other} types of pulsars related to NS--NS formation (Kalogera \& Lorimer
2000, hereafter KL00).
Progenitors of NS--NS systems experience two supernova explosions. 
The second supernova explosion (forming the NS that is
{\em not} observed as a pulsar) provides a unique tool for the study of
NS--NS formation, since the post-supernova evolution of the system is
simple, driven only by gravitational-wave radiation. There are three
possible outcomes after the second supernova: (i) a coalescing NS--NS is
formed (CB), (ii) a wide NS--NS (with a coalescence time longer than the
Hubble time) is formed (WB), or (iii) the binary is disrupted (D) and a
single pulsar similar to the ones seen in NS--NS systems is ejected.
Based on supernova orbital dynamics we can accurately calculate the
probability branching ratios for these three outcomes, $P_{\rm CB}$,
$P_{\rm WB}$, and $P_{\rm D}$. For a given kick magnitude, we can
calculate the maximum ratio $(P_{\rm CB}/P_{\rm D})^{\rm max}$ for the
complete range of pre-supernova parameters defined by the necessary
constraint $P_{\rm CB}\neq 0$ (Figure 1b). Given that the two types of
systems have a common parent progenitor population, the ratio of
probabilities is equal to the ratio of the birth rates $(BR_{\rm
CB}/BR_{\rm D})$.

We can then use (i) the absolute maximum of the probability ratio
($\approx 0.26$ from Figure 1b) and (ii) an empirical estimate of the
birth rate of single pulsars similar to those in NS--NS based on the
current observed sample to obtain an upper limit to the coalescence rate.
The selection of this sample involves some subtleties (KL00), and the
analysis results in $BR_{\rm CB} < 1.5\times 10^{-5}$\,yr$^{-1}$ (KL00).
Note that this number could be increased because of the small-number
sample and luminosity bias affecting this time the empirical estimate of
$BR_{\rm D}$ by a factor of $2-6$ (Kalogera et al.\ 1999).

This is an example of how we can use observed systems other than NS--NS to
improve our understanding of their coalescence rate. A similar calculation
can be done using the wide NS--NS systems instead of the single pulsars
(see KL00).

\section{Conclusions}

The current theoretical estimates of NS--NS coalescence rates appear to
have a rather limited predictive power. They cover a range in excess of 3
orders of magnitude and most importantly this range includes the value of
$10^{-5}$\,yr$^{-1}$ required for an ``enhanced'' LIGO detection rate of
2--3 events per year. This means that at the two edges of the range the
conclusion swings from no detection to many per month, and therefore the
detection prospects of NS--NS coalescence cannot be assessed firmly. On
the other hand empirical estimates derived based on the observed sample
appear to be more robust (estimates are all within a factor smaller than
100).  Given these we would expect a detection of one event every few
years up to ten events per year.

Estimates of the coalescence rate of BH--NS and BH--BH systems rely solely
on our theoretical understanding of their formation. As in the case of
NS--NS binaries, the model uncertainties are significant and the ranges
extend to more than 2 orders of magnitude. However, the requirement on the
Galactic rate is less stringent for 10\,M$_\odot$ BH--BH binaries, only
$\sim 2\times 10^{-7}$\,yr$^{-1}$. Therefore, even with the pessimistic
estimates for BH--BH coalescence rates ($\sim 10^{-7}$\,yr$^{-1}$), we
would expect at least a few detections per year, which is quite
encouraging. We note that a very recent examination of dynamical BH--BH
formation in globular clusters (Portegies-Zwart \& McMillan 1999) leads to
detection rates as high as a few per day.

 \acknowledgements
 I would like to thank the organizers of the meeting for their invitation
and support of my participation.  I also acknowledge support by the
Smithsonian Institute in the form of a CfA Post-doctoral Fellowship.

\end{document}